\documentclass[prb,twocolumn,superscriptaddress]{revtex4}
\usepackage{graphicx}
\usepackage{amsmath}
\usepackage{amsfonts}
\usepackage{amssymb,mathrsfs}

\newcommand{\bs}[1]{\boldsymbol{#1}}

\begin{document}

\title{Nonmonotonic magnetoresistance of a two-dimensional viscous
  electron-hole fluid in a confined geometry}

\author{P.S. Alekseev}
\affiliation{A.F. Ioffe Physico-Technical Institute, 194021 St. Petersburg, Russia}
\author{A.P. Dmitriev}
\affiliation{A.F. Ioffe Physico-Technical Institute, 194021 St. Petersburg, Russia}
\author{I.V. Gornyi}
\affiliation{Institut f\"ur Nanotechnologie,  Karlsruhe Institute of Technology,
76021 Karlsruhe, Germany}
\affiliation{\mbox{Institut f\"ur Theorie der kondensierten Materie,  Karlsruhe Institute of
Technology, 76128 Karlsruhe, Germany}}
\affiliation{A.F. Ioffe Physico-Technical Institute, 194021 St. Petersburg, Russia}
\affiliation{L.D. Landau Institute for Theoretical Physics, Kosygina street 2, 119334 Moscow, Russia}
\author{V.Yu. Kachorovskii}
\affiliation{A.F. Ioffe Physico-Technical Institute, 194021 St. Petersburg, Russia}
\affiliation{L.D. Landau Institute for Theoretical Physics, Kosygina street 2, 119334 Moscow, Russia}
\affiliation{Institut f\"ur Nanotechnologie, Karlsruhe Institute of Technology,
76021 Karlsruhe, Germany}
\author{B.N. Narozhny}
\affiliation{\mbox{Institut f\"ur Theorie der kondensierten Materie, Karlsruhe Institute of
Technology, 76128 Karlsruhe, Germany}}
\affiliation{National Research Nuclear University MEPhI (Moscow Engineering Physics Institute),
  115409 Moscow, Russia}
\author{M. Titov}
\affiliation{Radboud University Nijmegen, Institute for Molecules and Materials, NL-6525 AJ
Nijmegen, The Netherlands}
\affiliation{ITMO University, 197101 St. Petersburg, Russia}

\affiliation{}

\date{\today}

\begin{abstract}
   Ultra-pure conductors may exhibit hydrodynamic transport where the
   collective motion of charge carriers resembles the flow of a
   viscous fluid. In a confined geometry (e.g., in ultra-high quality
   nanostructures) the electronic fluid assumes a Poiseuille-like
   flow. Applying an external magnetic field tends to diminish viscous
   effects leading to large negative magnetoresistance. In
   two-component systems near charge neutrality the hydrodynamic flow
   of charge carriers is strongly affected by the mutual friction
   between the two constituents. At low fields, the magnetoresistance
   is negative, however at high fields the interplay between
   electron-hole scattering, recombination, and viscosity results in a
   dramatic change of the flow profile: the magnetoresistance changes
   its sign and eventually becomes linear in very high fields. This
   novel non-monotonic magnetoresistance can be used as a fingerprint
   to detect viscous flow in two-component conducting systems.
\end{abstract}

\maketitle

The independent particle approximation \cite{fey,ash} has dominated
the solid state physics for nearly a century. While clearly successful
in describing most of the basic transport phenomena in metals and
semiconductors, this approach completely neglects Coulomb interaction
between charge carriers (the latter is frequently said to be justified
by the ``weakness'' of electron-electron interaction due to, e.g.,
screening or statistical effects). To be more specific, in many
conventional conductors the typical interaction length scale,
$\ell_{ee}$, is too long in comparison to other relevant scales in the
system. In particular, at low temperatures the dominant scattering
process is due to potential disorder and hence the shortest length
scale is the mean free path, ${\ell_{\rm dis}\ll\ell_{ee}}$, which
determines the residual Drude resistivity at ${T=0}$. At high
temperatures, the electron-phonon interaction dominates,
${\ell_{ph}\ll\ell_{ee}}$. If these two temperature regimes overlap,
then indeed (at least, away from any phase transitions) the role of
electron-electron interaction is reduced to small
corrections. However, if there exists a temperature window where
${\ell_{ee}\ll\ell_{\rm dis},\ell_{ph}}$, then in that case the
independent particle approximation is violated: the motion of charge
carriers becomes collective (or hydrodynamic) and hence transport
properties of the system are determined by interaction \cite{us1}.

Signatures of the hydrodynamic behavior have been observed in recent
experiments in graphene \cite{exg1,exg2,exg3} and palladium cobaltate
\cite{exp}. The effect of external magnetic field on electronic
transport in systems with ${\ell_{ee}\lesssim\ell_{\rm
    dis},\ell_{ph}}$ was studied in magnetotransport measurements in
ultra-high-mobility GaAs quantum wells \cite{exs1,exs2,exs3} and more
recently in the Weyl semimetal WP$_2$ \cite{exw} reporting, in both
cases, large negative magnetoresistance. A theoretical explanation of
that effect has been recently suggested in Ref.~\onlinecite{pal} (see
also Ref.~\onlinecite{moore}) on the basis of a hydrodynamic model
where the viscosity coefficients are functions of temperature and
magnetic field.

A detailed account of the history of magnetotransport measurements is
beyond the scope of this paper. Here we only stress the following well
known facts: at the single-particle level, there is no classical
magnetoresistance (MR) in single-band (or one-component) systems;
taking into account more than one band of carriers (e.g., in
semiconductors) leads to the MR that is typically positive, quadratic
(${\sim{B}^2}$) at low fields, and saturating at classically high
fields (i.e., for ${\omega_c\tau\gg1}$, where $\omega_c$ is
the cyclotron frequency and $\tau$ is the disorder mean free
time); in some particular cases, MR in strong fields does not saturate
and may exhibit linear field dependence; and finally, in the majority
of situations MR is positive, while the negative MR occurs only in
special circumstances.

Nonsaturating MR has received considerable attention in recent
literature. Large positive (and often linear) MR was reported in
graphene and topological insulators close to charge neutrality
\cite{exl1,exl2,exl3,exl4,exl5,exl6,exl7,exl8,exl9}, Bi$_2$Te$_3$
nanosheets \cite{exb}, a topological material LuPdBi \cite{exlup},
semimetals WTe \cite{exwt1,exwt2}, NbP \cite{exn}, LaBi \cite{exla},
Cd$_2$As$_3$ \cite{exd}, and multilayer graphene \cite{exmg}. At the
same time, negative MR was found in Weyl semimetals \cite{exw1,exw2},
Dirac semimetals \cite{exds}, a novel semimetal TaAs$_2$ without Dirac
dispersion \cite{exr1}, and at the LaAlO$_3$/SrTiO$_3$ interface
\cite{exr2}.

Observations of negative MR in Weyl semimetals have attracted
substantial interest due to theoretical suggestions
\cite{son,spi,bur,lds} that this effect could be a direct condensed
matter manifestation of the Adler-Bell-Jackiw chiral anomaly
\cite{adl,jac,nn}. However, the unexpected variety of materials
exhibiting negative MR
\cite{exs1,exs2,exs3,exw,exw1,exw2,exds,exr1,exr2} does not support
the idea that such measurements may provide a ``smoking gun'' for
detecting a Weyl semimetal \cite{bur,lds}. Rather, there may be
several different mechanisms of negative MR similarly to the case of
linear positive MR that can appear, e.g., due to disorder
\cite{exl6,pl}, in the extreme quantum limit \cite{exlup,ab,Klier}, or
in compensated two-component systems
\cite{exl7,exl9,usg,us2,us3,sem}. In particular, the chiral anomaly
manifests itself in the negative longitudinal MR
\cite{exw2,son,spi,bur,lds} (i.e., the case of parallel electric and
magnetic fields, ${\bs{B}\|\bs{E}}$), while Ref.~\onlinecite{exw}
reports negative transverse MR.

In this paper, we consider the effect of the perpendicular magnetic
field on a two-dimensional (2D), two-component system of charge
carriers in the hydrodynamic regime at charge neutrality. While the
most obvious experimental realization of such system is monolayer
graphene, we consider the simplest case of two parabolic bands. We
believe that our qualitative results are independent of the particular
form of the spectrum and hence our theory is equally applicable to
bilayer graphene, topological insulators, or topological semimetals.

We show that in narrow rectangular samples (with the length much
larger than the width, ${L\gg{W}}$) a viscous electronic fluid
exhibits a Poiseuille-like flow. The spatial profile of the current
density in this flow is strongly affected by quasiparticle
recombination, impurity scattering, and mutual friction between the
two constituent subsystems (i.e., electron-hole scattering).

In the widest (in theoretical terms -- infinite) samples, transport
properties of the system are dominated by the impurity and (in the
two-component case) electron-hole scattering. Both processes have a
very similar effect leading to a finite Drude-like resistance. In
narrower samples, viscous effects start playing a role. As a result,
the flow becomes nonuniform. The typical scale of the spatial
variation of the current density, $\ell_G$, comprises the viscosity
and Drude mean free time. In the limit where this length exceeds the
sample width, ${\ell_G\gg{W}}$, we recover the standard parabolic flow
profile \cite{dau6,poi}. The corresponding resistance is then
proportional to the viscosity. This is a manifestation of the Gurzhi
effect \cite{gur} and therefore we refer to this length scale as the
Gurzhi length. In our theory, the Gurzhi resistance exceeds the
inviscid Drude resistance. Note, that this is not a contradiction to
either the recent observation of super-ballistic flow in graphene
\cite{exg3,lev} or the original Gurzhi effect \cite{gur}. The reason
is that we are considering the electronic fluid in the hydrodynamic
regime to begin with and do not compare it with a ballistic
(Knudsen-like \cite{knu}) regime where the resistance is determined by
the scattering off the system boundaries or large (macroscopic)
obstacles.

In the samples wider than the Gurzhi length, ${W\gg\ell_G}$, the flow
profile is modified: instead of the standard parabola we obtain a
variant of the catenary curve. Significant changes of the current
density are limited to a boundary layer of the size $\ell_G$, while
deep in the bulk the flow is nearly uniform, as if in the infinite
sample.

\begin{figure}[t]
\centerline{\includegraphics[width=0.55\columnwidth]{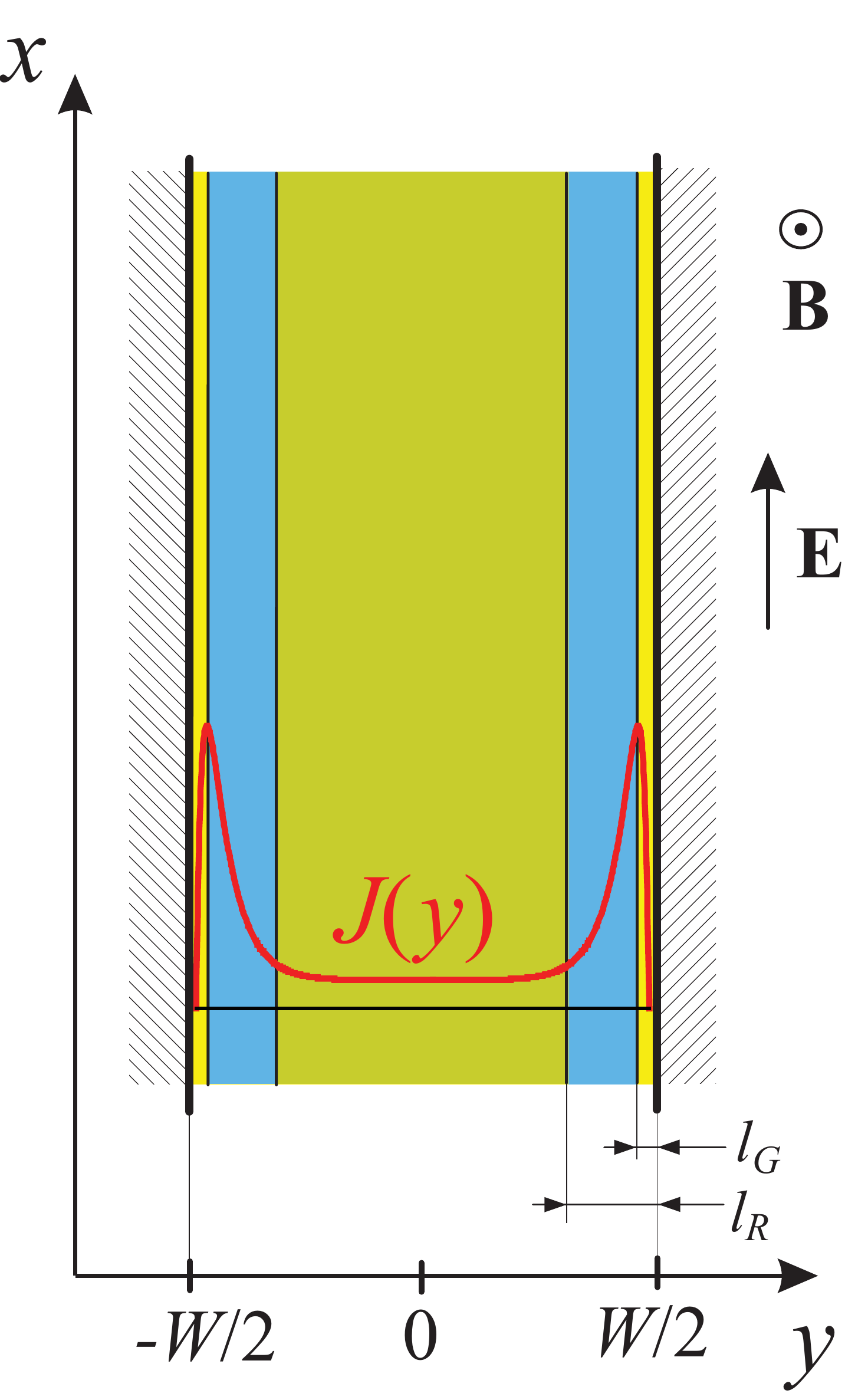}}
\caption{Inhomogeneous flow profile resulting from the interplay of
  electron-hole recombination, viscosity, impurity scattering, and
  mutual friction between the two system constituents. The profile is
  given for the case ${W\gg\ell_R\gg\ell_G}$.}
\label{fig:prof}
\end{figure}

Finally, the electron-hole recombination tends to create its own
boundary layer \cite{us2,us3,meg}. In this paper we consider the case
of weak (or slow) recombination, such that the corresponding length
scale is longer than the Gurzhi length, ${\ell_R>\ell_G}$. In the
case where both length scales are much smaller than the sample width,
the ``recombination layer'' is separated from the boundary by the
smaller ``Gurzhi layer'', see Fig.~\ref{fig:prof}. These layers are
characterized by a strongly inhomogeneous flow, in contrast to the
bulk of the system where the flow is uniform.

In the inviscid fluid, quasiparticle recombination processes strongly
affect the transport properties of the system leading to nonsaturating
(at charge neutrality), linear positive MR in strong fields
\cite{usg,us2,us3}. At the same time, viscous one-component systems
are characterized by negative MR \cite{exw,pal}. Here we show that
neutral two-component systems may exhibit both types of behavior at
the same time so that the MR is {\it non-monotonous}: in weak magnetic
fields, we find the negative, parabolic MR, while in strong fields the
resistance grows with the field eventually approaching the linear
dependence.

Our results are relevant to a wide range of novel materials including
compensated semimetals \cite{exw}, topological insulators, and
multilayer graphenes. Our main qualitative conclusions are independent
of the details of the quasiparticle spectrum and are applicable also
to systems with the linear (Dirac) spectrum such as the monolayer
graphene. As we specifically target the hydrodynamic, viscous flow of
charge carriers, we implicitly assume the regime of relatively high
temperatures (more precisely, our theory is justified in the
``hydrodynamic'' temperature window \cite{us1} with
${\ell_{ee}\ll\ell_{\rm{dis}},\ell_{ph}}$). Under this assumption, all
low-temperature quantum effects are washed out. In particular, Landau
quantization plays no role and we may consider very large magnetic
fields without running into quantum Hall physics.

\section{Compensated semimetals in two dimensional strip geometry}
\label{mtt}

Consider a two-component conductor (e.g., a narrow-band semiconductor
or a semimetal) in 2D. Allowing for electron-hole recombination, the
continuity equations for the two constituent densities can be written
as
\begin{equation}
\label{ceq}
\frac{\partial \delta n_{\alpha}}{\partial t}
+ \bs{\nabla}\cdot\bs{j}_{\alpha}
 = -\frac{ \delta n_{e}+\delta n_{h}}{2\tau_R},
\end{equation}
where $\alpha=e,h$ distinguishes the type of carriers, $\delta
n_\alpha$ are the deviations of the carrier densities from their
equilibrium values $n_\alpha^{(0)}$, $\bs{j}_{\alpha}$ are the carrier
currents, and $\tau_R$ is the electron-hole recombination time.

Charge transport in such systems can be described by a set of
macroscopic equations that are typically obtained by integrating the
kinetic (Boltzmann) equation \cite{pal,moore,us3}. In the
disorder-dominated regime such a theory can be reduced to a
generalized Ohm's law. In contrast, in the collision-dominated -- or
{\it hydrodynamic} -- regime the resulting macroscopic theory is a
generalization of the standard Navier-Stokes equation \cite{dau6}. In
the simplest case of two symmetric parabolic bands \cite{fn1}, the
continuity equations for the momentum densities -- or currents -- of
the two types of charge carriers are given by
\begin{eqnarray}
\label{eq0}
&&
\frac{\partial\bs{j}_\alpha}{\partial t}
+
\frac{\langle v^2 \rangle}{2}\bs{\nabla}\delta n_{\alpha}
-
\frac{e_\alpha n_\alpha^{(0)}}{m} \bs{E}
-
\omega_\alpha \left[ \bs{j}_\alpha \times \bs{e}_z \right]
=
\\
&&
\nonumber\\
&&
\qquad
=
-\frac{\bs{j}_\alpha}{\tau}
-\frac{\bs{j}_\alpha - \bs{j}_{\alpha'}}{2\tau_{eh}}
+
\eta_{xx} \Delta \bs{j}_\alpha
+
\eta_{xy}^\alpha \left[ \Delta \bs{j}_\alpha\times \bs{e}_z \right].
\nonumber
\end{eqnarray}
Here we consider the orthogonal magnetic field,
${\bs{B}\!=\!B\bs{e}_z}$; the electron and hole charges are
${e_h\!=\!e>0}$, ${e_e\!=\!-e}$, and the cyclotron frequencies are
${\omega_\alpha\!=\!e_\alpha{B}/(mc)\!=\!\omega_ce_\alpha/e}$; the
index $\alpha'$ denotes the constituent other than $\alpha$:
${\alpha'\!=\!e}$ for ${\alpha\!=\!h}$ and vice versa; $\tau_{eh}$ is
the momentum relaxation time due to electron-hole scattering; and the
averaging (for the parabolic spectrum with the constant density of
states $\nu_0$) is defined as \cite{usg}
\[
\langle\dots\rangle \!=\!-\!\!\int\!\!d\epsilon \, 
\frac{\partial f^{(0)}(\epsilon)}{\partial\epsilon} (\dots),
\]
where $f^{(0)}(\epsilon)$ is the Fermi distribution function. The two
viscosities, $\eta_{ij}$, depend on the magnetic field
\cite{pal,st}
\begin{equation}
\label{eta_xx}
\eta_{xx} \!=\! \eta_0/(1\!+\!4\omega_c^2 \tau_{ee}^2),
\quad
\eta_{xy}^\alpha \!=\! 2\omega_\alpha\tau_{ee}\eta_{xx},
\end{equation}
where $\eta_0$ is the shear viscosity in the absence of the magnetic
field 
\begin{equation}
\label{eta0}
\eta_0 \!=\! \langle v^4\rangle\tau_{ee}/(4\langle v^2\rangle)
\sim \langle v^2 \rangle \tau_{ee},
\end{equation}
and $\tau_{ee}$ is the electron-electron scattering time. 

Our hydrodynamic approach is justified if the electron-electron
scattering time $\tau_{ee}$ is the shortest time scale in the problem
(including the ``ballistic'' time defined by the sample width)
\begin{equation}
\label{cond}
\tau_{ee}\ll \tau, \tau_R, \tau_{eh}, \tau_B, \qquad \tau_B\sim W/\sqrt{\langle v^2\rangle}.
\end{equation}
In this case, the equations (\ref{eq0}) describe the two (electron and
hole) fluids that are weakly coupled by electron-hole scattering
\cite{us2,us3}. Unlike the single-component fluid considered in
Ref.~\onlinecite{pal}, these two fluids cannot be considered as
incompressible. However, under the assumption (\ref{cond})
electron-hole recombination dominates the viscous compressibility
(related to bulk viscosity) allowing us to drop the latter from
Eqs.~(\ref{eq0}). In the inviscid case \cite{usg,us2}, the
recombination-induced compressibility leads to positive linear
magnetoresistance in classically strong fields.

At charge neutrality, the currents and densities of the two
constituents are not independent. Introducing the total quasiparticle
density, ${\rho\!=\!n_e\!+\!n_h}$ and the linear combinations of the
two currents, ${\bs{P}\!=\!\bs{j}_e\!+\!\bs{j}_h}$ and
${\bs{j}\!=\!\bs{j}_h\!-\!\bs{j}_e}$, we re-write the hydrodynamic
theory (\ref{ceq}) and (\ref{eq0}) as
\begin{subequations}
\label{heqs}
\begin{equation}
\label{ceq1}
\frac{\partial\delta\rho}{\partial t} + \bs{\nabla}\cdot\bs{P}
 = -\frac{\delta\rho}{\tau_R},
\qquad
\bs{\nabla}\cdot\bs{j}=0,
\end{equation}
\begin{eqnarray}
\label{eqp}
&&
\frac{\partial\bs{P}}{\partial t}
+
\frac{\langle v^2 \rangle}{2}\bs{\nabla}\delta\rho
-
\omega_c \left[ \bs{j} \times \bs{e}_z \right]
=
-\frac{\bs{P}}{\tau}
+
\eta_{xx} \Delta\bs{P}
\\
&&
\nonumber\\
&&
\qquad\qquad\qquad\qquad\qquad\qquad\qquad\qquad
+
\eta_{xy} \left[ \Delta \bs{j}\times \bs{e}_z \right].
\nonumber
\end{eqnarray}
\begin{eqnarray}
\label{eqj}
&&
\frac{\partial\bs{j}}{\partial t}
-
\frac{e\rho^{(0)}}{m} \bs{E}
-
\omega_c \left[ \bs{P} \times \bs{e}_z \right]
=
\\
&&
\nonumber\\
&&
\qquad
=
-\frac{\bs{j}}{\tau}
-\frac{\bs{j}}{\tau_{eh}}
+
\eta_{xx} \Delta \bs{j}
+
\eta_{xy} \left[ \Delta \bs{P}\times \bs{e}_z \right].
\nonumber
\end{eqnarray}
\end{subequations}
Here ${\eta_{xy}\!=\!\eta_{xy}^\alpha{e}_\alpha/e}$ and
${\rho\!=\!\rho^{(0)}\!\!+\!\delta\rho}$ with
${\rho^{(0)}\!=\!n_e^{(0)}\!\!+\!n_h^{(0)}}$.

Finally, we consider the strip geometry, i.e., narrow rectangular
samples with the length, $L$, much larger than the width, $W$. In this
case, all physical quantities are functions of the transverse
coordinate $y$. At charge neutrality, the total electric field is
equal to the applied field, ${\bs{E}\!=\!(E,0)}$, which we assume to
be homogeneous. Requiring that no current flows out of the sides of
the sample, ${j_y(\pm{W}/2)\!=\!P_y(\pm{W}/2)\!=\!0}$, we find that
the electric current is directed along the strip,
${\bs{J}\!=\!e\bs{j}\!=\!e(j(y),0)}$, while the total quasiparticle
flow, ${\bs{P}\!=\!(0,P(y))}$, is orthogonal. Now we recall that in
the hydrodynamic regime the macroscopic currents vary on length scales
$\xi$ that are much longer than any other length scale in the
problem. Estimating the ratio of the Hall viscosity terms in
Eqs.~(\ref{heqs}) to the Lorentz terms as
${\eta_{xx}\tau_{ee}/\xi^2\ll1}$, we neglect the former and arrive at
the steady state equations
\begin{subequations}
\label{heqs2}
\begin{equation}
\label{ceq2}
P' = -\delta\rho/\tau_R,
\end{equation}
\begin{eqnarray}
\label{eqp2}
&&
\langle v^2 \rangle\delta\rho'/2
+
\omega_c j
=
-P/\tau
+
\eta_{xx} P''
\end{eqnarray}
\begin{eqnarray}
\label{eqj2}
&&
-
e\rho^{(0)} E/m
-
\omega_c P
=
-j/\tau
-j/\tau_{eh}
+
\eta_{xx} j''
\end{eqnarray}
\end{subequations}
The above equations describe the linear response transport in the
system in the hydrodynamic regime. Below we analyze their solutions
and discuss the applicability of the results.

\section{Results}

\subsection{Infinite sample}

Consider first the simplest case of an infinite sample. Here the
currents are uniform and the solution to Eqs.~(\ref{heqs2}) is
trivial:
\begin{subequations}
\label{ris}
\begin{equation}
\label{risjP}
j \!=\! j_0\! =\! \frac{\sigma_0 E}{1+\omega_c^2\tau\tau_*},
\quad
P \!=\! - \omega_c\tau j_0\!=\! - \frac{\sigma_0 E \omega_c\tau}{1+\omega_c^2\tau\tau_*},
\end{equation}
where
\begin{equation}
\label{s0}
\sigma_0 = e\rho^{(0)}\tau_*/m, \qquad \tau_* = \tau\tau_{eh}/(\tau+\tau_{eh}).
\end{equation}
\end{subequations}
In the absence of magnetic field only the longitudinal electric
current is flowing through the sample. The resistance is provided by
the impurity scattering and mutual friction between the electron and
hole subsystems. Once the field is applied, the lateral neutral
quasiparticle flow appears. In the clean limit,
${\tau\gg\tau_{eh},1/\omega_c}$, the longitudinal electric current
vanishes. If some disorder is present, then the result (\ref{risjP})
describes {\it positive} magnetoresistance.

\subsection{Finite-size sample without recombination}

Now we consider a sample of finite width, but neglect the
recombination processes. Then the continuity equation (\ref{ceq2})
combined with the hard-wall boundary conditions,
${P(\pm{W}/2)\!=\!0}$, yield the vanishing lateral flow, ${P\!=\!0}$,
while the equation (\ref{eqj2}) for the current $j$ has the form
\begin{equation}
\label{eqj3}
\ell_G^2(B) j''-j+\sigma_0E=0,
\end{equation}
where $\sigma_0$ is defined in Eq.~(\ref{s0}) and we have introduced
the Gurzhi length [field-dependent by means of Eq.~(\ref{eta_xx})].
\begin{equation}
\label{lg}
\ell_G(B) = \sqrt{\eta_{xx}\tau_*}
= \sqrt{\eta_0\tau_*}/\sqrt{1\!+\!(2\omega_c \tau_{ee})^2}.
\end{equation}
Assuming the standard hydrodynamic no-slip boundary conditions,
${j(\pm{W}/2)\!=\!0}$, we find the catenary profile
\begin{subequations}
\label{r1}
\begin{equation}
\label{rj1}
j = \sigma_0 E \left[1-\frac{\cosh y/\ell_G(B)}{\cosh W/[\ell_G(B)]}\right],
\end{equation}
with the total sample resistance
\begin{equation}
\label{rr1}
R = \frac{R_0}{1-\frac{2\ell_G(B)}{W}\tanh \frac{W}{2\ell_G(B)}}, \qquad
R_0 = \frac{L}{e\sigma_0 W}.
\end{equation}
\end{subequations}
Here $R_0$ is the sample resistance in the inviscid limit.

Assuming the large Gurzhi length, ${\ell_G(B)\gg{W}}$, we may expand the
expression for the electric current (\ref{rj1}) and recover the
parabolic profile typical of the standard Poiseuille flow
\cite{dau6,poi}. In this case, the sample resistance is proportional
to the sheer viscosity (a manifestation of the Gurzhi effect
\cite{gur})
\begin{equation}
\label{rg}
R\approx R_0 \frac{3\ell_G^2(B)}{W^2} 
=\frac{3Lm\eta_0}{e^2\rho^{(0)}W^3}\frac{1}{1\!+\!(2\omega_c \tau_{ee})^2}.
\end{equation}

\begin{figure}[t]
\centerline{\includegraphics[width=0.8\columnwidth]{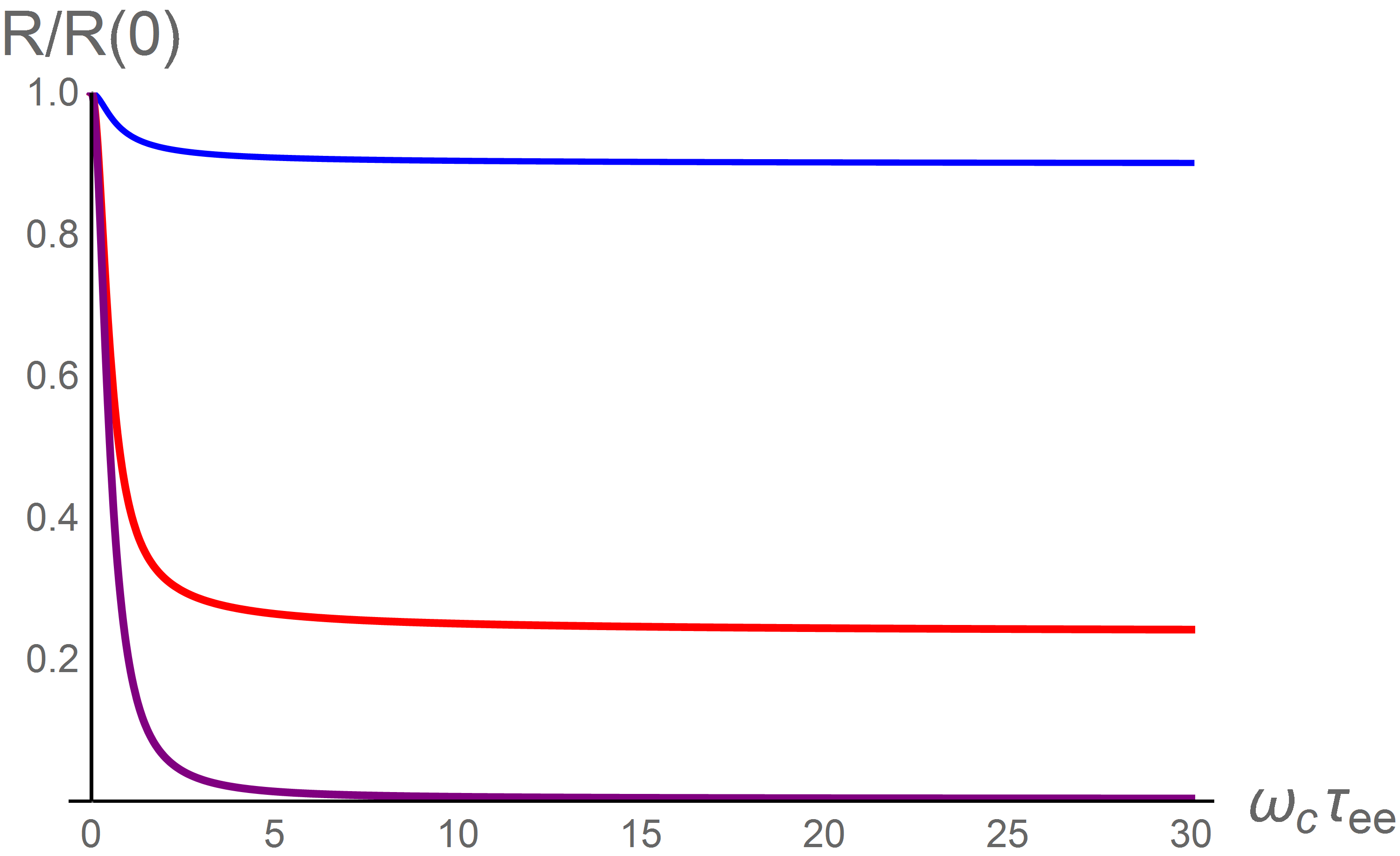}}
\caption{Resistance in the absence of recombination,
  Eq.~(\ref{rr1}). Different curves (top to bottom) correspond to
  different values of the ratio $W/[\ell_G(0)]=20,2,0.2$.}
\label{fig:r1}
\end{figure}

The resistance (\ref{rr1}) depends on the magnetic field only via the
field-dependent viscosity (\ref{eta_xx}). As the field is increased,
the viscosity -- and hence the Gurzhi length (\ref{lg}) -- decreases,
leading to {\it negative} magnetoresistance, see Fig.~\ref{fig:r1}. In
the case of a one-component fluid,
i.e. ${\tau_{eh}\rightarrow\infty}$, this effect was discussed in
Refs.~\onlinecite{pal,moore}.

\subsection{Finite-size sample with recombination}

Taking into account electron-hole recombination, we arrive at the
following equations
\begin{subequations}
\label{heqs4}
\begin{equation}
\label{eqj4}
\ell_G^2(B) j''-j+\sigma_0E+\omega_c\tau_*P=0,
\end{equation}
\begin{equation}
\label{eqp4}
\ell_R^2 P''-P-\omega_c\tau j=0,
\end{equation}
where the length scale describing the recombination processes is
\begin{equation}
\label{lr}
\ell_R\!=\!\sqrt{(\eta_{xx}\!+\!\langle v^2\rangle\tau_R/2)\tau}
\!\approx\!\sqrt{\langle v^2\rangle\tau_R\tau/2}\!\approx\! \ell_R(0).
\end{equation}
\end{subequations}
The length $\ell_R$ does in principle depend on the magnetic field
through the field-dependent viscosity, but in the limit of slow
recombination this contribution is neglected and the remaining
expression does not depend on $B$.

In the absence of the magnetic field the equations (\ref{heqs4})
decouple and we recover our previous results, ${P\!=\!0}$ and
Eq.~(\ref{r1}). In the presence of the magnetic field, the equations
(\ref{heqs4}) allow for a formal solution
\begin{subequations}
\label{r2}
\begin{equation}
\label{rjp2}
\begin{pmatrix}
j \cr
P
\end{pmatrix}
\!=\!
\left[ 1 \!-\!
\cosh (\widehat{M}^{\frac{1}{2}}y)\!\left[\cosh(\widehat{M}^{\frac{1}{2}}W/2)\right]^{-1}\right]\!\!
\begin{pmatrix}
j_0 \cr
-\omega_c\tau j_0
\end{pmatrix}\!,
\end{equation}
where the matrix $\widehat{M}$ is given by
\begin{equation}
\label{m}
\widehat{M} =
\begin{pmatrix}
\ell_G^{-2}(B) & -\omega_c\tau_*\ell_G^{-2}(B) \cr
\omega_c\tau\ell_R^{-2}(0) & \ell_R^{-2}(0)
\end{pmatrix}.
\end{equation}
The spatial variation of the currents is governed by the eigenvalues
of the matrix (\ref{m})
\begin{eqnarray}
\label{lambda}
&&
\lambda_\pm \!=\!
\frac{1}{2}\left[\ell_G^{-2}(B)\!+\!\ell_R^{-2}(0)\right] \!\pm\!
\\
&&
\nonumber\\
&&
\qquad
\pm
\sqrt{\!\left[\ell_G^{-2}(B)\!-\!\ell_R^{-2}(0)\right]^2\!/4\!
-\!\ell_G^{-2}(B)\ell_R^{-2}(0)\omega_c^2\tau\tau_*}
\!.
\nonumber
\end{eqnarray}
\begin{widetext}
Using the eigenvalues (\ref{lambda}), we express the current $j$ as
\begin{equation}
  \label{rj2}
  j \!=\! \frac{j_0}{\lambda_+\!-\!\lambda_-}
  \left[\left(1\!-\!\frac{\cosh \sqrt{\lambda_+}y}{\cosh \sqrt{\lambda_+}W/2}\right)
    \!\left[\ell_G^{-2}(B)(1\!+\!\omega_c^2\tau\tau_*)\!-\!\lambda_-\right]
    \!-\!\left(1\!-\!\frac{\cosh \sqrt{\lambda_-}y}{\cosh \sqrt{\lambda_-}W/2}\right)
    \!\left[\ell_G^{-2}(B)(1\!+\!\omega_c^2\tau\tau_*)\!-\!\lambda_+\right]
    \right]\!.
\end{equation}
This leads to the following expression for the resistance of the
sample
\begin{eqnarray}
  \label{rr2}
  &&
  R=R_0(\lambda_+\!-\!\lambda_-)
  \left[
    \left(1\!-\!\frac{2}{\sqrt{\lambda_+}W}\tanh \frac{\sqrt{\lambda_+}W}{2}\right)
    \left(\ell_G^{-2}(B)\!-\!\frac{\lambda_-}{1\!+\!\omega_c^2\tau\tau_*}\right)
  \right.
  \\
  &&
  \nonumber\\
  &&
  \qquad\qquad\qquad\qquad\qquad\qquad\qquad\qquad
  \left.
    -
    \left(1\!-\!\frac{2}{\sqrt{\lambda_-}W}\tanh \frac{\sqrt{\lambda_-}W}{2}\right)
    \left(\ell_G^{-2}(B)\!-\!\frac{\lambda_+}{1\!+\!\omega_c^2\tau\tau_*}\right)
    \right]^{-1}.
  \nonumber
\end{eqnarray}
\end{widetext}
\end{subequations}

The general expressions (\ref{r2}) can be simplified using the general
assumption (\ref{cond}). Indeed, the hydrodynamic approach holds if
the electron-electron scattering time $\tau_{ee}$ is much shorter than
any other time scale, including the recombination time. Hence, the
Gurzhi length (\ref{lg}) is much smaller than the recombination length
(\ref{lr}):
\[
\tau_R\gg\tau_{ee} \quad\Rightarrow\quad
\ell_R(0)\gg\ell_G(0).
\]
This allows us to expand the square root in Eq.~(\ref{lambda}). In
this paper, we are interested in the case where the eigenvalues
(\ref{lambda}) are real (in general, them may become complex leading
to an interesting oscillatory behavior of the currents; this effect
will be discussed elsewhere). The necessary condition justifying this
assumption is
\[
\ell_R(0)/\sqrt{2\!+\!4\omega_c^2\tau\tau_*}>\ell_G(B) 
\quad{\rm or}\quad \tau_R\gtrsim\tau_*^2/\tau_{ee}.
\]
For simplicity, we will now assume a stronger inequality,
\begin{equation}
  \label{sass}
  \tau_R\gg\tau_*^2/\tau_{ee}.
\end{equation}
The latter assumption simplifies the algebra, but does not lead to a
qualitative change in the results.

Expanding the square root in Eq.~(\ref{lambda}), we find
\begin{equation}
  \label{mapp}
  \lambda_+\approx\ell_G^{-2}(B), \qquad
  \lambda_-\approx\frac{1\!+\!\omega_c^2\tau\tau_*}{\ell_R^2(0)}\equiv\ell^{-2}_R(B).
\end{equation}
Under the assumptions (\ref{mapp}) and (\ref{sass}), the resistance
(\ref{rr2}) simplifies and we arrive at the result
\begin{eqnarray}
  \label{rr3}
  &&
  R=R_0
  \left[
    \left(1\!-\!\frac{2\ell_G(B)}{W}\tanh \frac{W}{2\ell_G(B)}\right)
  \right.
  \\
  &&
  \nonumber\\
  &&
  \qquad\qquad
  \left.
    -
    \left(1\!-\!\frac{2\ell_R(B)}{W}\tanh \frac{W}{2\ell_R(B)}\right)
    \frac{\omega_c^2\tau\tau_*}{1\!+\!\omega_c^2\tau\tau_*}
    \right]^{-1}\!.
  \nonumber
\end{eqnarray}
Depending on the parameter values, the expressions (\ref{rr2}) and
(\ref{rr3}) may exhibit positive, negative, or nonmonotonic
magnetoresistance. Below we discuss the emerging parameter regimes.

\section{Discussion}

The following discussion of the field dependence of the sample
resistance (\ref{rr3}) is based on the simple properties of the
function
\[
f(z)=1-\frac{\tanh z}{z}\rightarrow
\begin{cases}
  z^2/3, & z\ll 1, \cr
  1-1/z, & z\gg 1.
\end{cases}
\]

\subsection{Finite-size sample without recombination}

In the absence of recombination (i.e., for ${\ell_R(0)\!\rightarrow\!\infty}$),
we may neglect the second term in Eq.~(\ref{rr3}) and hence recover
Eq.~(\ref{rr1}).

Using the explicit expression for the viscosity coefficient
(\ref{eta_xx}), we re-write the ratio ${\ell_G(B)/W}$ as
\[
\frac{\ell_G(B)}{W} = \frac{\ell_G(0)}{W}\frac{1}{\sqrt{1+4\omega_c^2\tau_{ee}^2}}.
\]
Substituting this expression in Eq.~(\ref{rr1}), we find that the
magnetoresistance is always {\it negative} and can be {\it very large}
in strong fields, ${R_{{\rm{max}}}/R_{{\rm{min}}}\gg1}$.

\begin{figure}[b]
\centerline{\includegraphics[width=0.8\columnwidth]{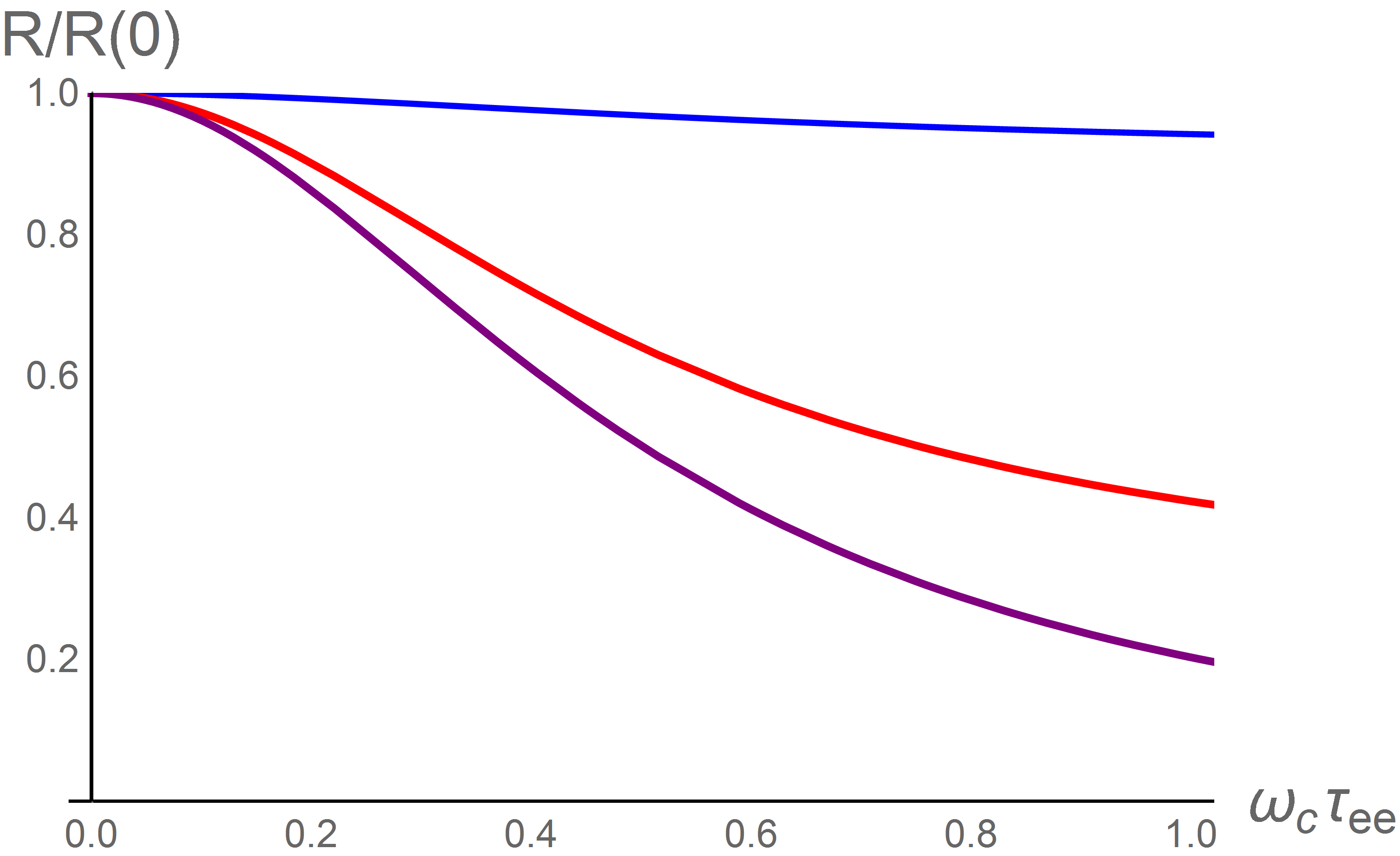}}
\caption{Resistance in the absence of recombination, Eq.~(\ref{rr1})
  in weak magnetic fields. Different curves (top to bottom) correspond
  to different values of the ratio $W/[\ell_G(0)]=20,2,0.2$.}
\label{fig:r2}
\end{figure}

The resulting field dependence is shown in Figs.~\ref{fig:r1} and
\ref{fig:r2}. While the qualitative features of the MR are independent
of the particular value of the ratio ${W/\ell_G(0)}$, the effect is
much more pronounced in narrow samples.

In weak fields, the MR is quadratic
\[
\frac{R(B)}{R(0)}\approx 1-2
\left[\frac{\tanh^2\frac{W}{2\ell_G(0)}}
{1\!-\!\frac{2\ell_G(0)}{W}\tanh\frac{W}{2\ell_G(0)}}\!-\!1\right]
\omega_c^2\tau_{ee}^2.
\]
In very strong fields,
${\omega_c\tau_{ee},\omega_c\tau_{ee}W/\ell_G(0)\gg1}$, the resistance
saturates approaching the inviscid value $R_0$.

The behavior of the system in the absence of recombination is similar
to that of the one-component system discussed in
Ref.~\onlinecite{pal}. Strictly speaking, the one-component limit is
achieved neglecting electron-hole scattering, i.e.,
${\tau_{eh}\gg\tau}$. In this case, the electron and hole subsystems
are independent and our system consists of two copies of the system
discussed in Ref.~\onlinecite{pal}. At the same time, electron-hole
scattering and impurity scattering affect the electric current in the
same way. Indeed, the corresponding time scales, $\tau_{eh}$ and
$\tau$ enter the equation (\ref{eqj2}) in a symmetric fashion. As a
result, in the absence of recombination the effect of electron-hole
scattering is reduced to renormalizing the transport mean free time,
$\tau_*$.

\subsection{Finite-size sample with recombination}

In the presence of recombination, the magnetoresistance is determined
by the interplay between the sample width, $W$, and the two length
scales, ${\ell_R\gg\ell_G}$. The field dependence of the recombination
length $\ell_R$ is similar to that of the Gurzhi length, however, the
typical field scales are rather different:
\[
\frac{\ell_R(B)}{W} = \frac{\ell_R(0)}{W}\frac{1}{\sqrt{1+\omega_c^2\tau\tau_*}},
\]
where under our assumptions
\[
\tau\tau_*\gg\tau_{ee}^2.
\]

In the absence of the magnetic field, the sample resistance is
independent of the weak (under our assumptions) recombination. As a
result, $R(0)$ is the same as above. Furthermore, in the Gurzhi limit
(\ref{rg}) this value is independent of $\tau_{eh}$ as well.

\begin{figure}[t]
\centerline{\includegraphics[width=0.8\columnwidth]{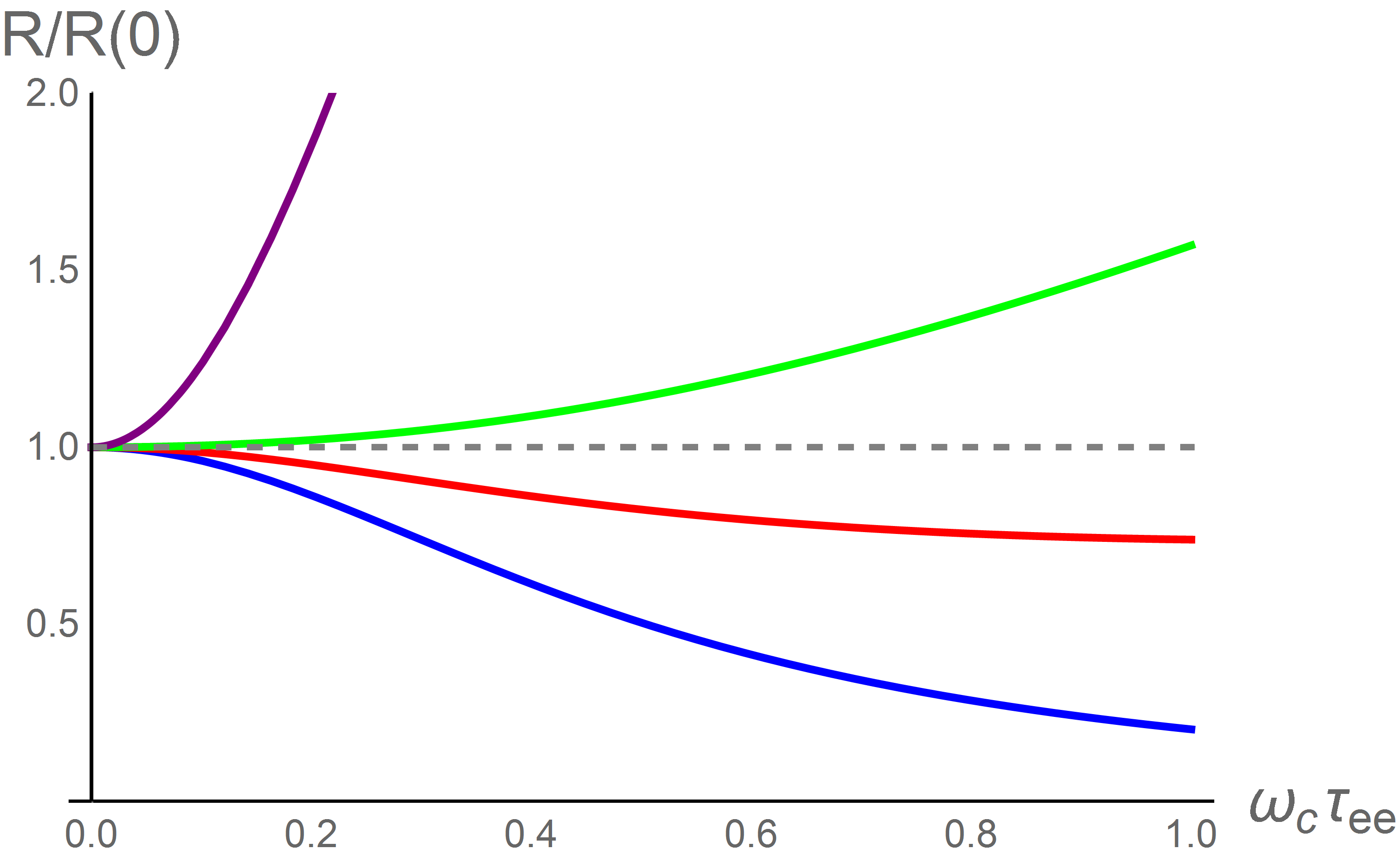}}
\caption{Resistance (\ref{rr3}) in weak magnetic fields. The four
  curves (top to bottom: ${W/[\ell_G(0)]\!=\!80,12,4,0.1}$) illustrate
  the four parameter regimes in Eq.~(\ref{a1}). The numerical values
  correspond to the choice $\ell_R(0)/\ell_G(0)=40$,
  $\tau\tau_*/\tau_{ee}^2=100$.}
\label{fig:r4}
\end{figure}

\subsubsection{Weak fields}

In weak fields, ${B\rightarrow0}$, the MR is still quadratic,
\begin{equation}
\label{rw}
\frac{R(B)}{R(0)}\approx 1 - A_1 \omega_c^2\tau_{ee}^2,
\end{equation}
but with the coefficient that depends on $\tau_R$, $\tau$, and $\tau_*$
\begin{eqnarray*}
&&
A_1=
2\left[\frac{\tanh^2\frac{W}{2\ell_G(0)}}
{1\!-\!\frac{2\ell_G(0)}{W}\tanh\frac{W}{2\ell_G(0)}}\!-\!1\right]
\\
&&
\\
&&
\qquad\qquad\qquad\qquad
-
\left[1\!-\!\frac{2\ell_R(0)}{W}\tanh\frac{W}{2\ell_R(0)}\right]\frac{\tau\tau_*}{\tau_{ee}^2}.
\end{eqnarray*}

For the narrowest samples, $W\ll\ell_G(0)$, the coefficient $A_1$ is
determined by the electron-electron scattering time
\[
A_1(W\rightarrow0)\approx 4,
\]
leading to {\it negative} MR.

\begin{figure*}[t]
\centerline{\includegraphics[width=0.8\textwidth]{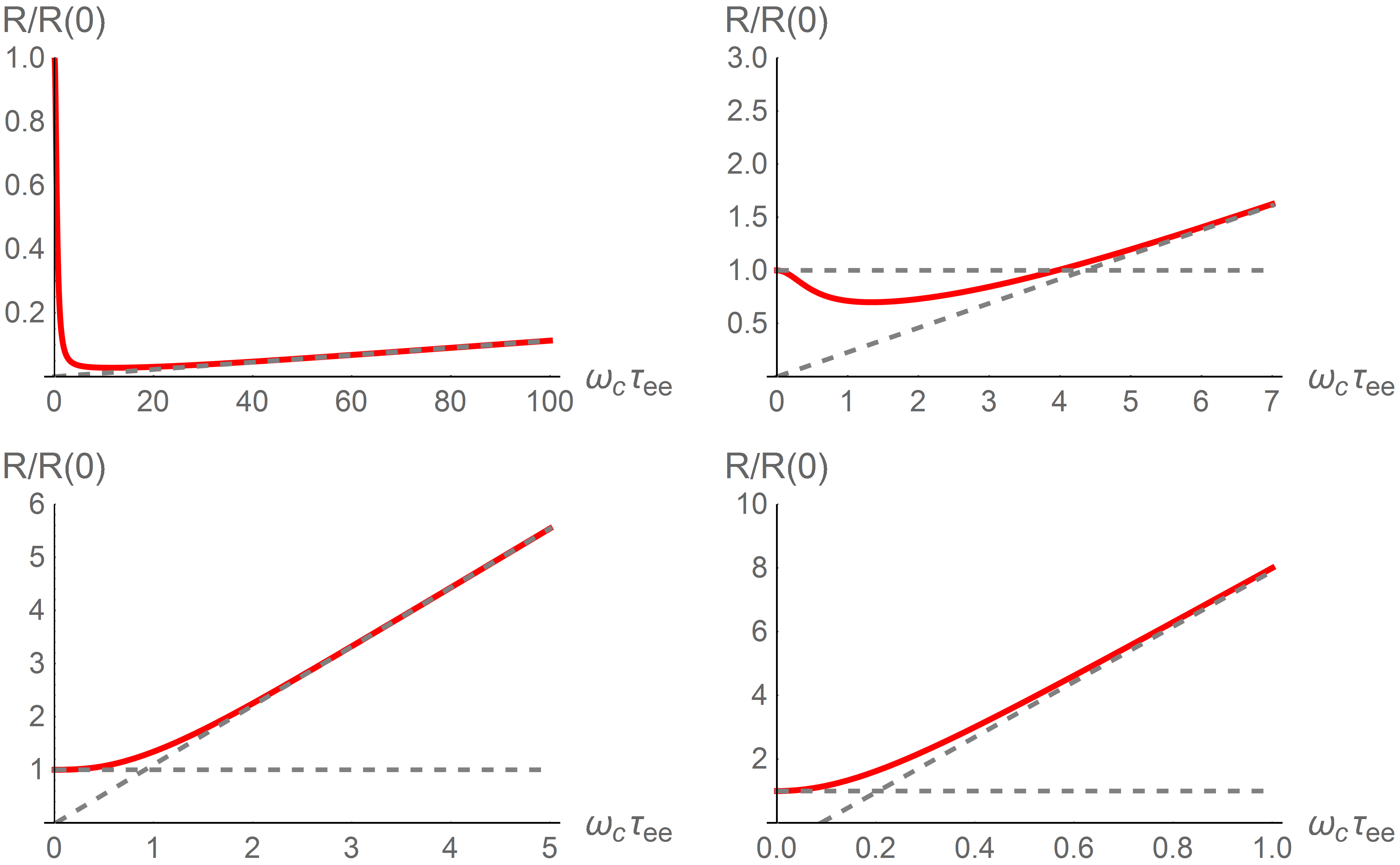}}
\caption{The four regimes of Eq.~(\ref{a1}) illustrated in the four
  panels (clockwise, from top left:
  ${W/[\ell_G(0)]\!=\!0.5,4,12,80}$). The numerical values correspond
  to the choice ${\ell_R(0)/\ell_G(0)\!=\!50}$,
  ${\tau\tau_*/\tau_{ee}^2\!=\!100}$. The dashed line describes the
  strong-field limit (\ref{a2}).}
\label{fig:r41}
\end{figure*}

For wider samples the coefficient $A_1$ changes sign. This sign change
occurs when
\begin{equation}
\label{W0}
W= W_0, \qquad W_0^3 \simeq 48\frac{\tau_{ee}^2}{\tau\tau_*}\ell_R^2(0)\ell_G(0).
\end{equation}
Hence, for $W<W_0$, the MR is still negative, albeit with a smaller
coefficient, while for $W>W_0$ the MR becomes {\it positive}. For the
widest samples, $W\gg\ell_R(0)$, we find strong positive MR. The four
parameter regimes can be summarized as
\begin{equation}
  \label{a1}
A_1\approx
\begin{cases}
  4, & W\ll\ell_G(0), \cr
  4\ell_G(0)/W, & \ell_G(0)\ll W < W_0, \cr
  -\dfrac{W^2}{12\ell_R^2(0)} \dfrac{\tau\tau_*}{\tau_{ee}^2}, & W_0<W\ll\ell_R(0), \cr
  -\tau\tau_*\tau_{ee}^{-2}, & W\gg\ell_R(0).
\end{cases}
\end{equation}
The weak-field magnetoresistance (\ref{rw}) is illustrated in
Figs.~\ref{fig:r4} and \ref{fig:r41}.

\subsubsection{Strong fields}

In strong fields, ${B\rightarrow\infty}$, we recover {\it positive}
linear MR \cite{usg,us2,us3}. This behavior corresponds to the
following regime of parameters: ${\omega_c\tau_{ee}\gg1}$ and
$\omega_c^2W^2\tau_*\gg\tau_R\langle{v}^2\rangle$. The resulting
resistance is given by
\begin{equation}
  \label{a2}
  \frac{R(B)}{R_0} \approx A_2
  \left[\omega_c\tau_{ee} - A_2\frac{\tau_{ee}^2}{\tau\tau_*}\right],
\end{equation}
where
\[
A_2 = \frac{W}
{2\ell_R(0)\frac{\tau_{ee}}{\sqrt{\tau\tau_*}}-\ell_G(0)}.
\]
Here the denominator remains positive as long as the eigenvalues
(\ref{lambda}) are real, i.e. for
${\tau_R>\tau_*^2/\tau_{ee}}$. Consequently, in wide samples,
$W>\ell_R(0)$, the MR is always positive, see Fig.~\ref{fig:r41} for
illustration.

\subsubsection{Intermediate fields}

The behavior of the resistance (\ref{rr3}) in between the above two
asymptotic regimes is strongly affected by the sample width.

In the narrowest samples, ${W\ll\ell_G(0)}$, and not too strong
magnetic fields,
${\omega_c\tau_{ee}\ll\sqrt{\tau_*\tau_{ee}}/\tau_B}$, we recover the
Gurzhi limit, ${W\ll\ell_G(B)}$, with the Lorentzian-shaped resistance
given by Eq.~(\ref{rg}).  In stronger fields,
${\sqrt{\tau_*\tau_{ee}}/\tau_B\ll\omega_c\tau_{ee}\ll(\tau_{ee}/\tau_B)\sqrt{\tau_R/\tau_*}}$,
the width of the sample, ${\ell_G(B)\ll{W}\ll\ell_R(B)}$, enters the
intermediate parameter range. Here, the resistance 
\begin{equation}
\label{rin}
\frac{R(B)}{R_0}\approx 1 + \frac{\ell_G(0)}{\omega_c\tau_{ee}W} +
\frac{\omega_c^2\tau_{ee}W^2}{12\ell_G^2(0)}
\frac{\tau_*^2}{\tau_R},
\end{equation}
remains close to its minimum value
\begin{subequations}
\label{rmin}
\begin{equation}
\label{rm}
R_{\rm min} = R_0\left[1 + {\cal O}\left(\tau_*^{2/3}(\tau_R\tau_{ee})^{-1/3}\right)\right]
\approx R_0,
\end{equation}
that is achieved at
\begin{equation}
\label{bm}
\omega_c \rightarrow \omega_c^* = W_0/(2\tau_{ee}W).
\end{equation}
\end{subequations}
In the strongest fields,
${\omega_c\tau_{ee}\gg(\tau_{ee}/\tau_B)\sqrt{\tau_R/\tau_*}}$, the
recombination length becomes smaller than the sample width and we
recover the above linear MR.

\begin{figure*}[t]
\centerline{\includegraphics[width=0.95\textwidth]{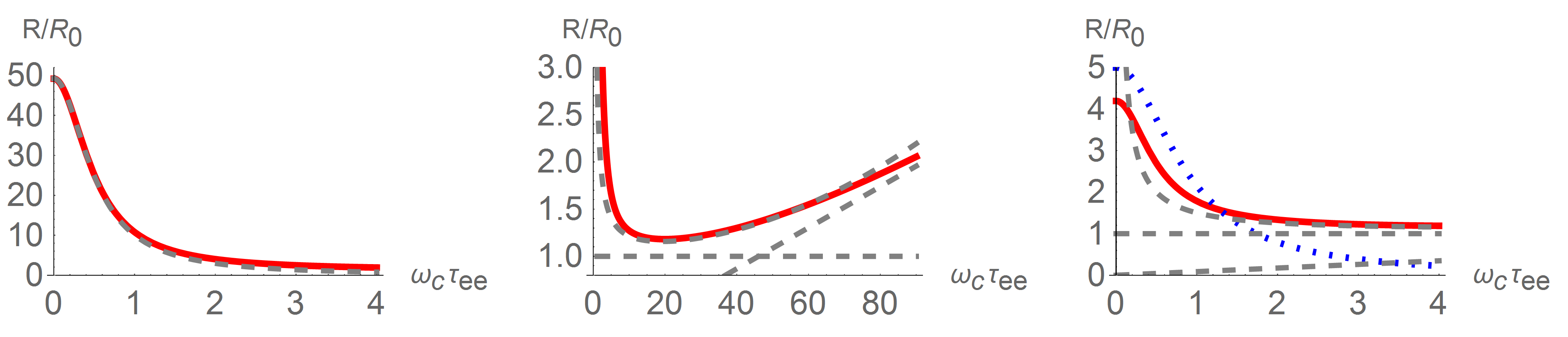}}
\caption{Intermediate regimes of the resistance. The left panel
  illustrates the Gurzhi limit in the narrowest samples,
  Eq.~(\ref{rg}), taking into account higher-order corrections in
  ${W/\ell_G(0)\!=\!0.5}$ while plotting the Lorentzian (dashed
  curve). The central panel shows Eq.~(\ref{rin}) for the same sample
  width (dashed curve). The numerical values were computer for
  ${\ell_R(0)/\ell_G(0)\!=\!120}$,
  ${\tau\tau_*/\tau_{ee}^2\!=\!100}$. The straight dashed line
  describes the strong-field limit (\ref{a2}). The right panel shows
  the onset of the intermediate regime (\ref{rin}) for a wider sample
  with ${W/\ell_G(0)\!=\!2}$. The dotted line shows the Lorentzian
  (\ref{rg}).}
\label{fig:r5}
\end{figure*}

In wider samples, the parameter windows of the above intermediate
regimes shrink and gradually disappear, see Fig.~\ref{fig:r5}. For
${\ell_G(0)\ll{W}\ll\ell_R(0)}$, the condition of the Gurzhi limit is
violated and the Lorentzian (\ref{rg}) is no longer a good
approximation, while the onset of the intermediate regime (\ref{rin})
is shifted towards weaker fields, see the right panel in
Fig.~\ref{fig:r5}. For the widest samples, ${W\gg\ell_R(0)}$, the
nonmonotonic behavior disappears and we find positive MR, see the two
bottom panels in Fig.~\ref{fig:r41}.

\begin{figure}[b]
\centerline{\includegraphics[width=0.8\columnwidth]{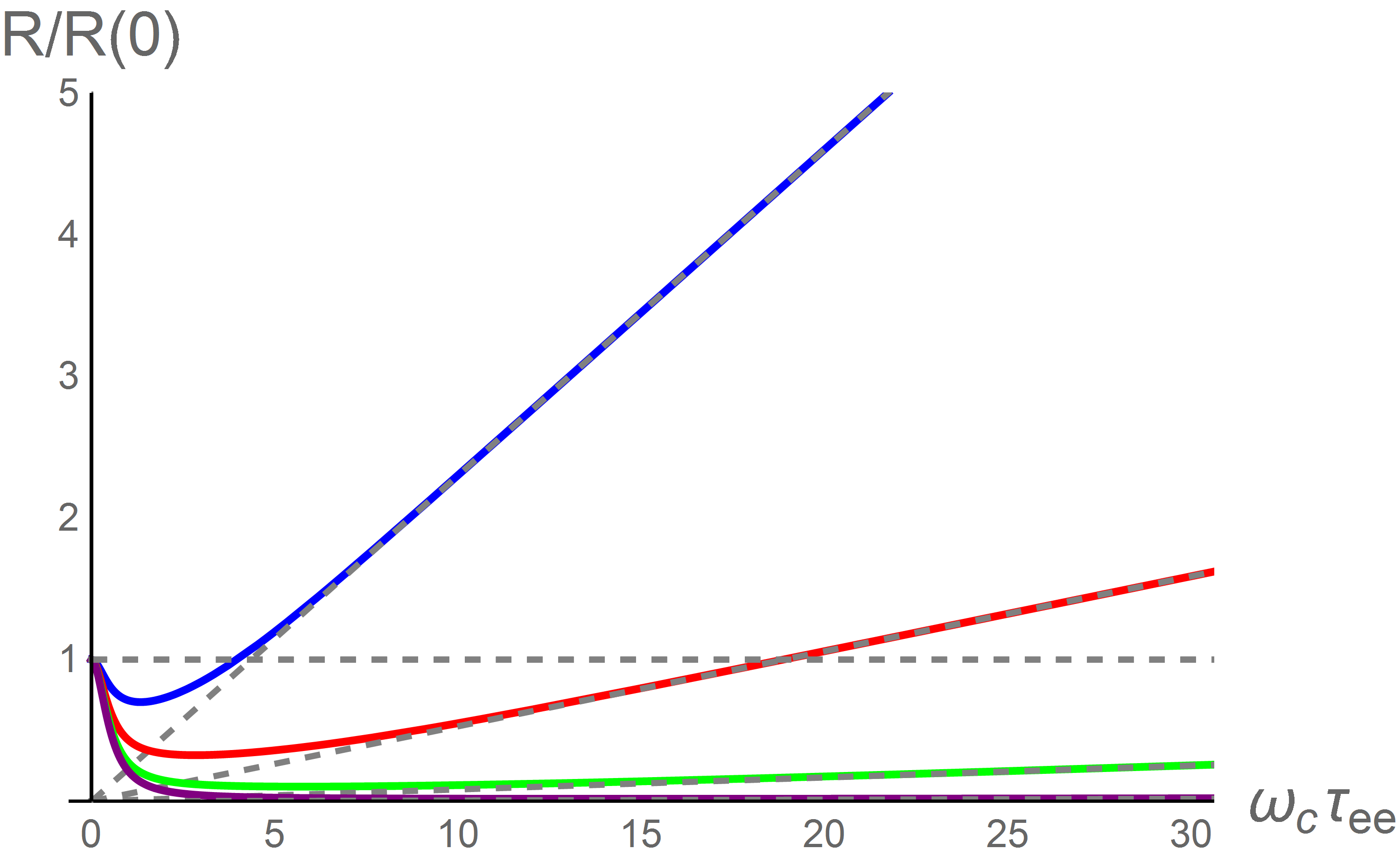}}
\caption{Nonmonotonic behavior of the resistance
  (\ref{rr3}). Different curves (top to bottom) correspond to
  different values of the ratio ${W/[\ell_G(0)]\!=\!4,2,1,0.4}$. The
  numerical values correspond to the choice $\ell_R(0)/\ell_G(0)=50$,
  $\tau\tau_*/\tau_{ee}^2=100$. Dashed lines describe the strong-field
  limit (\ref{a2}).}
\label{fig:r3}
\end{figure}

The {\it nonmonotonic} MR occurs in intermediate-width samples, $W <
W_0$: it is negative in weak fields, but becomes positive in strong
fields. The magnetoresistance (\ref{rr3}) is illustrated in
Figs.~\ref{fig:r41}~-~\ref{fig:r3}. In particular,
the sample resistance as a function of the magnetic field develops a
minimum, ${R_{\rm min}\approx R_0}$, at a particular value of the
field, $B^*$, see Eqs.~(\ref{rmin}).

\subsection{Phase diagram }

\begin{figure}[b]
\centerline{\includegraphics[width=1.0\linewidth]{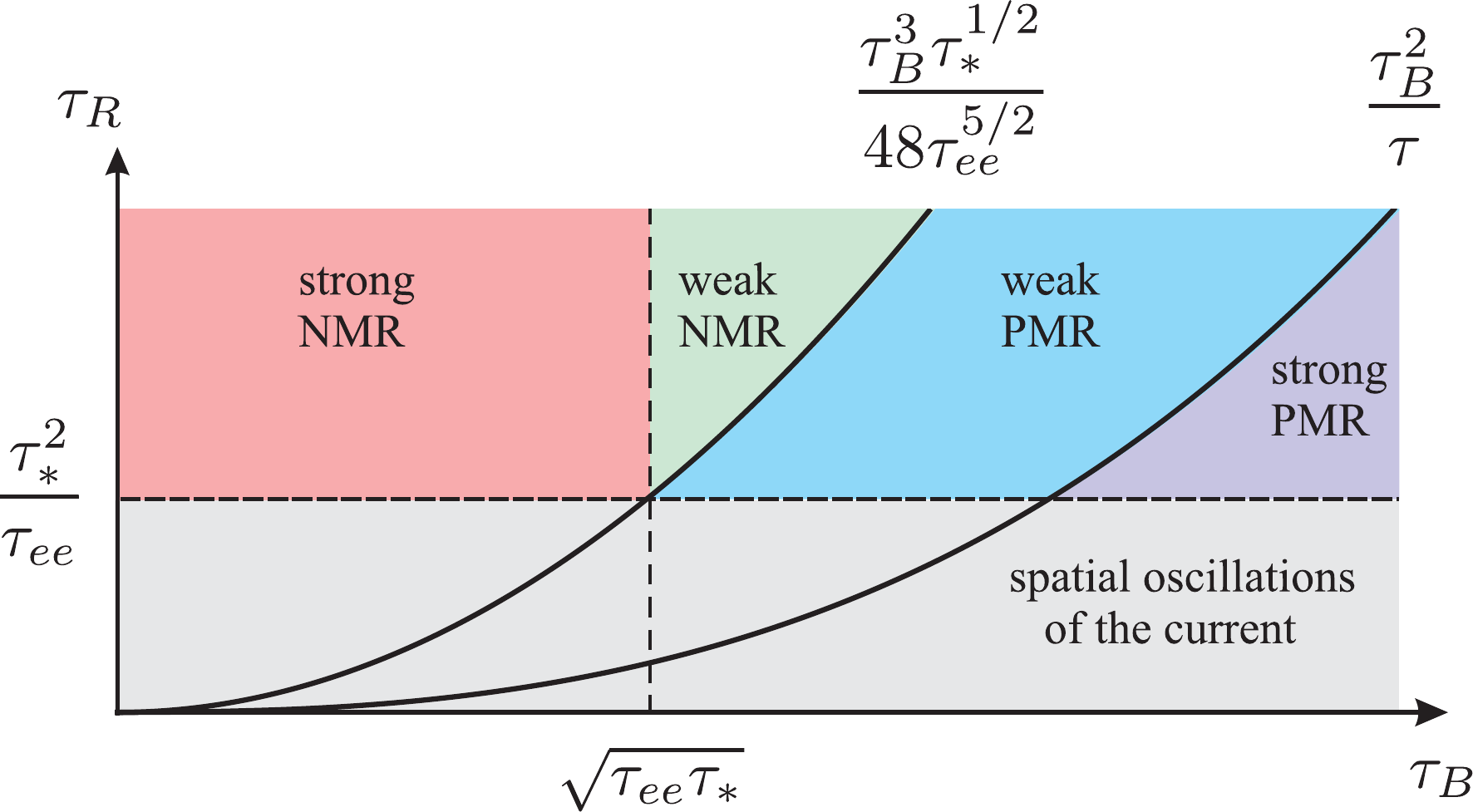}}
\caption{Summary of the parameter regimes exhibited by the sample
  resistance subjected to the external magnetic field,
  Eq.~(\ref{rr3}). The lower part of the diagram with
  ${\tau_R\ll\tau_*^2/\tau_{ee}}$ describes the regime where the
  eigenvalues (\ref{lambda}) are complex and the currents exhibit
  oscillatory behavior. This regime is beyond the scope of this paper
  and will be addressed elsewhere. The labels ``weak/strong NMR
  (PMR)'' refer to the negative (positive) MR in weak magnetic fields,
  see Eq.~(\ref{a1}).}
\label{Fig-phase}
\end{figure}

The expression (\ref{rr3}) for the sample resistance shows a rich
variety of parameter regimes illustrated in
Figs.~\ref{fig:r4}~-~\ref{fig:r5}. The conditions determining these
regimes can be summarized in a phase-diagram-like manner, see
Fig.~\ref{Fig-phase}, where they are shown in terms of the
recombination length, $\ell_R$, and the sample width represented by
the ``ballistic'' length,
${\tau_B\sim{W}/\sqrt{\langle{v}^2\rangle}}$, see
Eq.~(\ref{cond}). Under the assumption of the weak recombination
adopted in this paper, Eq.~(\ref{sass}), the eigenvalues
(\ref{lambda}) are real within the whole range of magnetic fields.
Varying the sample width with a fixed value of $\tau_R$, we may scan
through different MR regimes, see Fig.~\ref{Fig-phase}.

The narrowest samples are described by the condition
${W\ll\ell_G(0)}$, that can be re-written as
${\tau_B\ll\sqrt{\tau_{ee}\tau_*}}$. In this case, we observe strong
negative MR, see the top left panel in Fig.~\ref{fig:r41}. For wider
samples, ${\ell_G(0)\ll{W}<W_0}$, or
${\sqrt{\tau_{ee}\tau_*}\ll\tau_B\ll\tau_R^{1/3}\tau_{ee}^{5/6}\tau_*^{-1/6}}$,
the weak field MR is still negative, but is characterized by a small
coefficient $A_1$, see Eq.~(\ref{a1}). In Fig.~\ref{Fig-phase}, we
refer to this regime as ``weak negative MR''. Overall in this regime,
the resistance is a {\it non-monotonic} function of the field: in
stronger fields the recombination processes dominate and lead to
linear positive MR, see Fig.~\ref{fig:r3}.

Wider samples exhibit positive MR, see the bottom panels in
Fig.~\ref{fig:r41}. Since the coefficient $A_1$ increases drastically
as the width of the sample exceeds the zero-field recombination
length. Consequently, we refer to the regime
${\tau_B\gg\sqrt{\tau_R\tau}}$ as the regime of strong positive MR in
Fig.~\ref{Fig-phase}.

Note, that the latter regime disappears in the ultra-clean limit, where
formally ${\tau\rightarrow\infty}$. In this case, both eigenvalues
$\lambda_\pm$ remain real and finite, although the field-dependent
recombination length $\ell_R(B)$ defined by Eq.~(\ref{mapp}) becomes
inverse proportional to the magnetic field:
\[
\ell_R(B) \to (1/\omega_c) \sqrt{\langle v^2 \rangle \tau_R/(2 \tau_{eh})},
\]
where we have used the explicit form (\ref{lr}) of $\ell_R(0)$ with
${\tau_*\rightarrow\tau_{eh}}$. The resistance \eqref{rr3} simplifies
somewhat and becomes
\begin{equation}
\frac{R}{R_0} \!=\! 
\left[  
\frac{2}{\omega_c\tau_W} \tanh\frac{\omega_c\tau_W}{2}  
\!-\!    
\frac{2\ell_G(B)}{W} \tanh\frac{W}{2\ell_G(B)} 
\right]^{-1}\!\!,
\label{ultra}
\end{equation}
where
\[
\tau_W=\tau_B\sqrt{2\tau_{eh}/\tau_R}.
\]
In the regime of weak recombination,
${\tau_R\gg\tau_{eh}^2/\tau_{ee}}$, see Eq.~(\ref{sass}), the
expression \eqref{ultra} exhibits all three regimes shown in the phase
diagram, Fig.~\ref{Fig-phase}, to the left of the strong positive MR
regime.

\section{Conclusions}

We considered transport properties of a 2D viscous, two-component
electronic liquid at charge neutrality. We showed, that in the narrow
strip geometry, the fluid exhibits a Poiseuille-like inhomogeneous
flow where the spatial variation of the current density is controlled
by the quasiparticle recombination and viscous effects described by
the length scales $\ell_R$ and $\ell_G$, respectively, see
Fig.~\ref{fig:prof}. Assuming that the recombination length exceeds
the viscous Gurzhi length at all fields, ${\ell_R(B)\gg\ell_G(B)}$, we
find that the inhomogeneity is strongest near the sample edges.

The contribution of the boundary regions to transport coefficients is
strongly affected by the external magnetic field leading to strong,
nonmonotonous magnetoresistance. In weak fields, the MR is quadratic
in $B$, see Eq.~(\ref{rw}), and can be positive or negative with the
sign change occuring at a particular value of the sample width
(\ref{W0}). This behavior is illustrated in Figs.~\ref{fig:r4} and
\ref{Fig-phase}.  Depending on parameter values, we distinguish four
different regimes of strong and weak, positive and negative MR
summarized in Eq.~(\ref{a1}).

In strong fields, the MR is positive and linear in $B$
\cite{usg,us2,us3}, see Eq.~(\ref{a2}). In the parameter regimes where
the weak-field MR is negative, this leads to a nonmonotonic behavior
where the resistance curves exhibit a minimum (\ref{rmin}) at a
particular value of the field, $B^*$, see Figs.~\ref{fig:r4} and
\ref{fig:r3}.

In this paper we have considered the electronic system in the
hydrodynamic regime. Any temperature dependence of transport
coefficients appears by means of the temperature dependence of the
scattering times $\tau_{ee}$, $\tau_{eh}$, and $\tau_R$. This
temperature dependence is beyond the scope of the hydrodynamic
approach and has to be derived from a microscopic theory (see,
e.g. Ref.~\onlinecite{hyd} for such derivation in graphene). We have
assumed, see Eq.~(\ref{cond}), that intra-band scattering is more
effective than inter-band scattering, i.e. ${\tau_{eh}\gg\tau_{ee}}$.
While this can be easily achieved in double-layer electron-hole
structures \cite{drag}, we believe that at the qualitative level our
results are unaffected by this assumption and remain valid in a more
general case where ${\tau_{eh}\sim\tau_{ee}}$. We have also neglected
the effects of the Hall viscosity, see discussion in the text between
Eqs.~(\ref{heqs}) and (\ref{heqs2}). Indeed, the shortest length scale
describing spatial variation of the current is $\xi\sim\ell_G(B)$. In
this case, the smallness of the Hall viscosity terms (in comparison to
the Lorentz terms) in Eqs.~(\ref{heqs}) is justified by our main
assumption (\ref{cond}).

The effects described in this paper can be observed experimentally in
any 2D compensated semimetal, including graphene, topological
insulators, and narrow band semiconductors. Recently, the hydrodynamic
behavior has been observed in Weyl semimetals \cite{exw}. These
materials are three-dimensional and show the behavior discussed in
this paper in the case where the current is perpendicular to the
magnetic field \cite{exw}. In the inviscid limit, neutral 3D systems
with the slab geometry behave very similarly to 2D systems
\cite{us3}. We expect that our present results for viscous two-fluid
flows are applicable is this case as well.

\acknowledgments

We thank M.I. Dyakonov, A.D. Mirlin, D.G. Polyakov, J. Schmalian, and
M. Sch\"utt for fruitful discussions. This work was supported by the
Dutch Science Foundation NWO/FOM 13PR3118 (MT), the Russian Foundation
for Basic Research Grant 17-02-00217 (VYK), the Russian Science
Foundation Grant 17-12-01182 (PSA, APD, IVG), and the MEPhI Academic
Excellence Project, Contract No. 02.a03.21.0005 (BNN).


\begin{thebibliography}{99}
\bibitem{fn1}
  The case of linear spectrum in, e.g., graphene or 2D surface of 3D
  topological insulators will be considered elsewhere. At the
  qualitative level, however, we expect that the main conclusions of
  this paper will hold even in this special case.
\end{thebibliography}
\end{document}